\documentclass[aps,preprint,showpacs]{revtex4}
\usepackage{epsfig}
\usepackage{amssymb}
\usepackage{multirow}
\usepackage{dcolumn}
\usepackage{bm}
\usepackage[nottoc]{tocbibind}
\usepackage[ruled,vlined,onelanguage]{algorithm2e}

\usepackage{lineno,hyperref}
\modulolinenumbers[5]
\usepackage[utf8]{inputenc}
\usepackage{amsmath}
\modulolinenumbers[1]
\usepackage{siunitx}
\usepackage{amsmath}
\usepackage[capitalise]{cleveref}
\usepackage{siunitx}

\begin{document}

\title{Human Trafficking in Mexico: Data sources, Network Analysis and the Limits of Dismantling Strategies}

\author{Sof\'ia de la Mora Tostado$^1$}
\author{Mayra N\'u\~nez-L\'opez$^1$} 
\email{mayra.nunez@itam.mx}
\author{Esteban A. Hern\'andez-Vargas$^2$} 
\email{esteban@systemsmedicine.de}

\affiliation{$^1$ Department of Mathematics, Instituto Tecnol\'ogico Aut\'onomo de M\'exico, R\'io Hondo 1, 01080 Ciudad de M\'exico, Mexico}

\affiliation{$^2$ Institute of Mathematics, UNAM, Mexico}

\begin{abstract}
Human trafficking is a heartless crime that represents the second most profitable crime in the world.
Mexico's geographical position makes it a country with high levels of human trafficking.
Using the snowball sampling method, the major contribution of this paper is the abstraction of the human trafficking network on the southern border of Mexico. 
Based on a social network analysis, it is identified that the criminal network is moderately centralized (44.32\%) and with medium density (0.401).
Therefore, the network has minimal cohesiveness and members may find it difficult to share information, money, or products among themselves.
To evaluate different dismantling strategies to tackle the criminal organization, three algorithms are evaluated.
We found that the first actors to be removed are neither the most connected nor the most peripheral, 
but the actors who are moderately connected to people of their kind should be removed. 
In summary, this paper provides a significant step forward to understand quantitatively human trafficking networks and evaluate the limits of dismantling strategies. 
\end{abstract}

\maketitle

\section{Introduction}
Human trafficking is a ruthless crime that seeks to exploit people for different purposes. 
The International Labor Organization estimates that, in the world, there are 40.3 million people who are victims of this crime, which represents the second most profitable criminal activity in the world, after weapon trafficking \cite{ILO2014}. 
Human trafficking has persisted over the years because there is a market for exploiting individuals for affordable prices. As long as the social demand continues to exist, it will continue to be covered \cite{Casillas2011}. 

Mexico has one of the highest rates of human trafficking in the world \cite{Barrios2021}.
This can be attributed to institutional instability, discrimination, corruption, and other causes. There have been very few efforts taken to combat human trafficking in the country. The violence unleashed in Mexico in the twenty-first century to combat drug trafficking has demonstrated that tear-down techniques based on the elimination of random individuals were ineffective \cite{Dell2015}.

Human trafficking, as any other form of criminal activity, is a complex problem that requires multidisciplinary approaches for its effective prevention and combat. Social networks and the respective collection of data within criminology stemming have received a considerable interest in the scientific community to abstract illicit networks \cite{Bindu2017,Granados2021}. 
The actors in a criminal network, which are the nodes in a network, are most often people but occasionally organizations.
Actors are connected to each other as family, friends, or rivalries, these connections are represented by the edges in a network \cite{Faust2019}.
Studying how elements of a network are connected can help uncover salient patterns in the criminal network. The connections between data elements are often assumed to be observed, at least partially. However, in many real-world applications, the configuration interpretation is not given a priori and needs to be inferred from the available data.

Strategies for the disintegration of criminal networks have been largely proposed in the literature, for example, by deleting nodes randomly, \cite{Keegan2010}, or simply removing nodes with essential connections, \cite{Duijn2014a}. 
More complex techniques like the Social Network Analysis (SNA) can help find weak points in the criminal network, supporting a more thorough investigation and more efficient network application decision-making \cite{Duijn2014}. Another example is in terrorist networks, where it was found that removing nodes with a lot of connections and subsequently removing nodes with privileged roles were equally beneficial in diminishing the network \cite{Xu2003}. Similarly, the Cosa Nostra network was used to devise the most effective assault \cite{Musciotto2022}, network analysis determined that actors affiliated with the mafia syndicate were crucial in destroying the network. On the other hand, the criminal intelligence network \cite{Cunha2018} discovered that by removing just 2\% of the nodes the network could be dismantled.
Crime disruption strategies strongly depend on both network topology and network resilience \cite{Duijn2014a}. However, as criminal networks operate in secrecy, data-driven knowledge concerning the effectiveness of different criminal network disruption strategies is very limited. 

In this paper we are focused on Chiapas, which is one of Mexico's poorest states, located in the southest of the country, it has a border with Guatemala that stretches for 658 kilometers, \cite{Chiapas}. 
In Chiapas, there are three very common ways to attract a person for human trafficking. One example is foreigners from Central and South America, as well as a few Asians, Caribbeans, and Africans, who seek to go to Mexico in order to enter the United States of America. They arrive in Chiapas and search for jobs in the area while they figure out how to keep moving forward. Very commonly they turn to illicit nightlife companies because of their immigration status, where they end up being exploited for labor. 
Another case occurs when Central Americans are promised a decent job in Chiapas so that they can get out of the situation of poverty. They ask for little money to help them cross the border and promise them a stable life, yet these people are being recruited to exploit them. The last case occurs when someone in a vulnerable situation is approached by a person in a park, a nightclub, or a shopping center offering them work or a love relationship and ends up being sexually exploited \cite{Casillas2011}.

To abstract a human trafficking network of Chiapas, we developed a social network analysis by collecting real data with the snowball 
sampling methodology from a border state in the southern part of Mexico.
With the human traffic network, we performed a network analysis to further characterize and then understand the criminal organization in Tapachula, Chiapas. 
Consequently, we evaluate potential dismantling strategies and their respective limitations.
\section{Data Collection on the southern border of Mexico}
To construct the trafficking network of Tapachula we follow the snowball sampling method that locates network members using basic interviews \cite{Goodman1961}.
Given a population, a random sample is taken and each of the elements in the sample, in this case, each of the people in the sample, are asked to name $k$ individuals in the population, who share trafficking-related connections. The process continues with the new contacts who mention the people they know related to the crime.

Interviews were online with victims, traffickers, and ex-convicts of human trafficking in the Soconusco region. 
The interviews for the building of the network in question were conducted in the months of June, July, and August of 2020.
For the safety of authors, victims, interviewees, and the interviewer, the identities of the performers and their locations will not be stated; instead, everything will be referenced using keys.
The person who conducted and published the interviews was contacted and provided information on the interviewees. With the data of the interviewer, people who suffered from trafficking in the past or participated in the business were contacted. 
The general roles in a human trafficking organization described in \cite{Casillas2011} are as follows:
	
The \textit{Exploiters} manage the activities of the treaty. In turn, they can have the following roles: 
(\textit{i}) The \textit{Recruiters} are the ones who attract prey in public places, in Chiapas they look for migrants who are looking to reach a specific location and deceive them;
(\textit{ii}) The \textit{Raiteros} are dedicated to driving the trucks or taxis that move the treaties to the places where they are exploited every day.

The \textit{Correspondents} are in charge of transporting and taking the tracts to their place of work. These are subdivided:
(\textit{i}) The \textit{Escorts}, as their name implies, accompany the migrants from Guatemala to the safe house, they hire taxis or 
trucks to transport them and sometimes they are members of the police;
(\textit{ii}) The \textit{Guides} are in charge of passing the migrants across the border, sometimes they are called polleros.

The \textit{Collectors} collect money from traffickers who are trafficked and pay the traffickers.
The \textit{Couriers} are spies and they are dedicated to informing others about possible movements by the police.
The \textit{Minders} monitor the houses in which the treaties live, maintain order among the victims, and take care that the authorities do not approach their trading points or the victims.
\textit{Officials} are an essential part of the network. They forgive dealers in exchange for receiving some of their merchandise for free or at a special price.
\textit{Informants} are submerged in the government or the police, specifically, they are dedicated to passing valuable information to traffickers to avoid being captured.
	
The initial contact was made possible by non-governmental organizations dedicated to assisting victims in their rescue and reintegration into society. Some of these organizations also collaborate with reformed ex-convicts to gain insight into the industry and identify potential victims. The organizations that they helped with contacts work in various states around Mexico. Although they were limited to contact with the criminals and the company, the victims were able to witness the behavior of the traffickers and their relationships and played a significant part in understanding the network. Through phone calls, five survivors of human trafficking were questioned; from these interviews, useful information was obtained about the organization, but not about the criminals' personal contacts. It was feasible to reach five people who were still participating in the crime after interviewing three ex-convicts; two of them asked not to specify their involvement in the network, and the other three performed the roles of participant, raitero, and recruiter.

To rebuild the network as accurately as possible, questions regarding other persons participating were asked, as well as questions about the network’s activities and the roles that members performed. To do this, a list of generic questions was compiled using the established process. Five survivors, six ex-convicts, and eleven people still participating in the crime at the time were interviewed in the end. The interviewer orally asked each of the interviewees if they knew the people who mentioned them before. Only relationships that were validated by both actors involved are included. People who were victims and did not participate in the business are not included in the data. For the rest of the people found, a category was assigned according to the function they fulfilled within the network of human trafficking under study.

\section{Human trafficking network of the southern border of Mexico}
\begin{figure}[ht!]
	\includegraphics[scale=0.4]{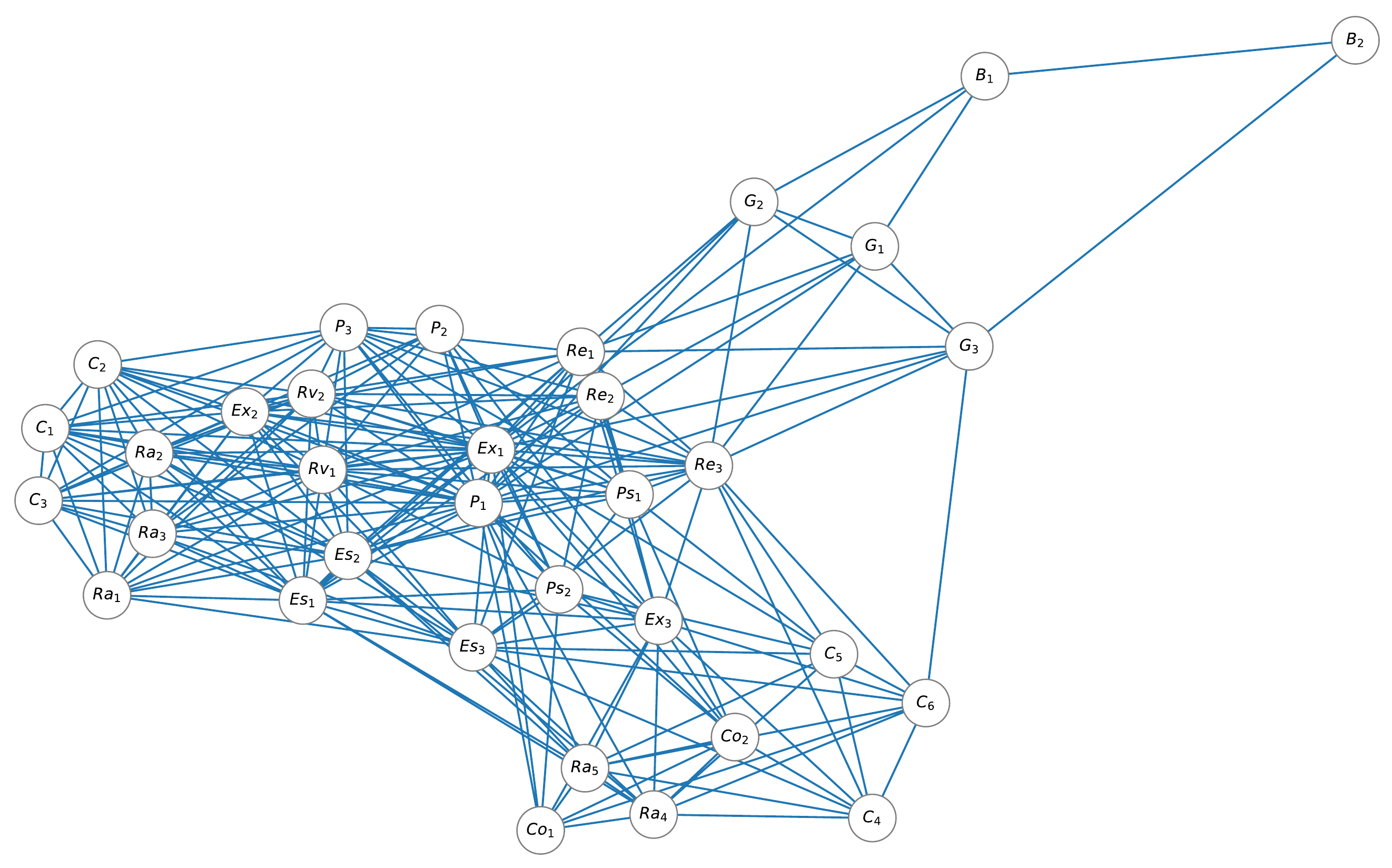}
	\caption{Human trafficking network in Chiapas, Mexico. The actors of the network (gray circles) are labelled as follows: 
		Caretaker ($C_i$), Company ($Co_i$), Body Guard ($B_i$), Estafeta ($Es_i$), Exploiter ($Ex_i$), Public Servant ($Ps_i$), Guide ($G_i$), Participant ($P_i$), Raitero ($Ra_i$), Recruiter ($Re_i$), and Recruiter/Victim ($Rv_i$). The subindex $i$ indicate the number of the actor in case exist more than one in that position. Blue lines represent the connections with the respective actors.}
	\label{fig:network}
\end{figure}

The snowball sampling methodology, explained in the previous section, was used to find the largest possible number of people involved in a human trafficking network in Chiapas.
After the data collection, the human trafficking network was built, see Figure \ref{fig:network}, this consists of 34 nodes (or actors) and 225 edges (or connections). The roles of each one of those involved are given in Table \ref{roles}.
Even though 34 actors and 225 links between them were discovered, this does not mean that the network is made up of just these elements, is common in a criminal network that a part of it is hard to discover \cite{Granados2021}

\begin{table}[ht]
	\begin{tabular}{c l l}
		\hline\hline
		Role  &  Definition \\ [0.1ex] 
		\hline
		Caretaker & Person in charge of maintaining order between the treaties and \\
		&  monitor theirs  results. & \\
		Company & Entity that “hires” people to exploit them.\\
		Body Guard & Person who hires transportation to cross migrants from Guatemala \\ &  to Mexico. & \\
		Estafeta & Inform others about possible operations that affect the  network. \\
		Exploiter & Manages trafficking activities in the network. \\
		Public Servant & Government worker who allows the networks to exist in exchange \\ 
		&  for money. & \\
		Guide & Colloquially called “pollero”. \\
		Participant & Police that allow trafficking and take advantage of it. \\
		Raitero & The driver of the network. \\
		Recruiter & The person who enrolls others in trafficking through deceit. \\
		 Recruiter/Victim  & The person who enrolls others in trafficking and was previously a\\
		 &  victim of the network. &  \\
		\hline\hline
	\end{tabular}
	\caption{Roles in the human trafficking network in Tapachula, Chiapas.}
	\label{roles}
\end{table}

\subsection{Social Network Analysis}
In this analysis, the components of a network are called \emph{actors} and \emph{links}, these are analyzed using a set of centrality metrics to identify the most important actors in the network. We considered the usual metric in networks science like density, diameter, betweenness, clustering coefficient, among others.

In the context of this work, the \textit{clustering coefficient} is the proportion of all paths in the network that are closed and allows finding groups or gangs into which a network is divided. Without technicalities, the clustering coefficient of each node indicates how close they are to a gang and the \textit{average clustering coefficient} is the average of the clustering coefficients of all nodes. This may be thought of as the typical behavior of nodes in the vicinity of each other \cite{Everton2012}. Given that the purpose of this research is to defragment a network and quantify the fragmentation, a method for recognizing unconnected pairs of nodes is described. A measure of fragmentation may be defined as follows, given the adjacency matrix $A$ with entries $A _{i,j}$, where $i$ and $j$ are real values and $n$ is the number of nodes:
\\
\begin{equation}
	F=1-\frac{2\sum\limits_{i}\sum\limits_{j<i}A_{i,j}}{n(n-1)}
	\label{EqF}
\end{equation}
\\
where $F=1$ if no node is connected to another and $F=0$ if all nodes are connected to each other, that is, it is a complete graph \cite{Borgatti2002}.

\begin{figure}[ht]
   \begin{center}
    \footnotesize{(a)}\\
    \includegraphics[scale=0.30]{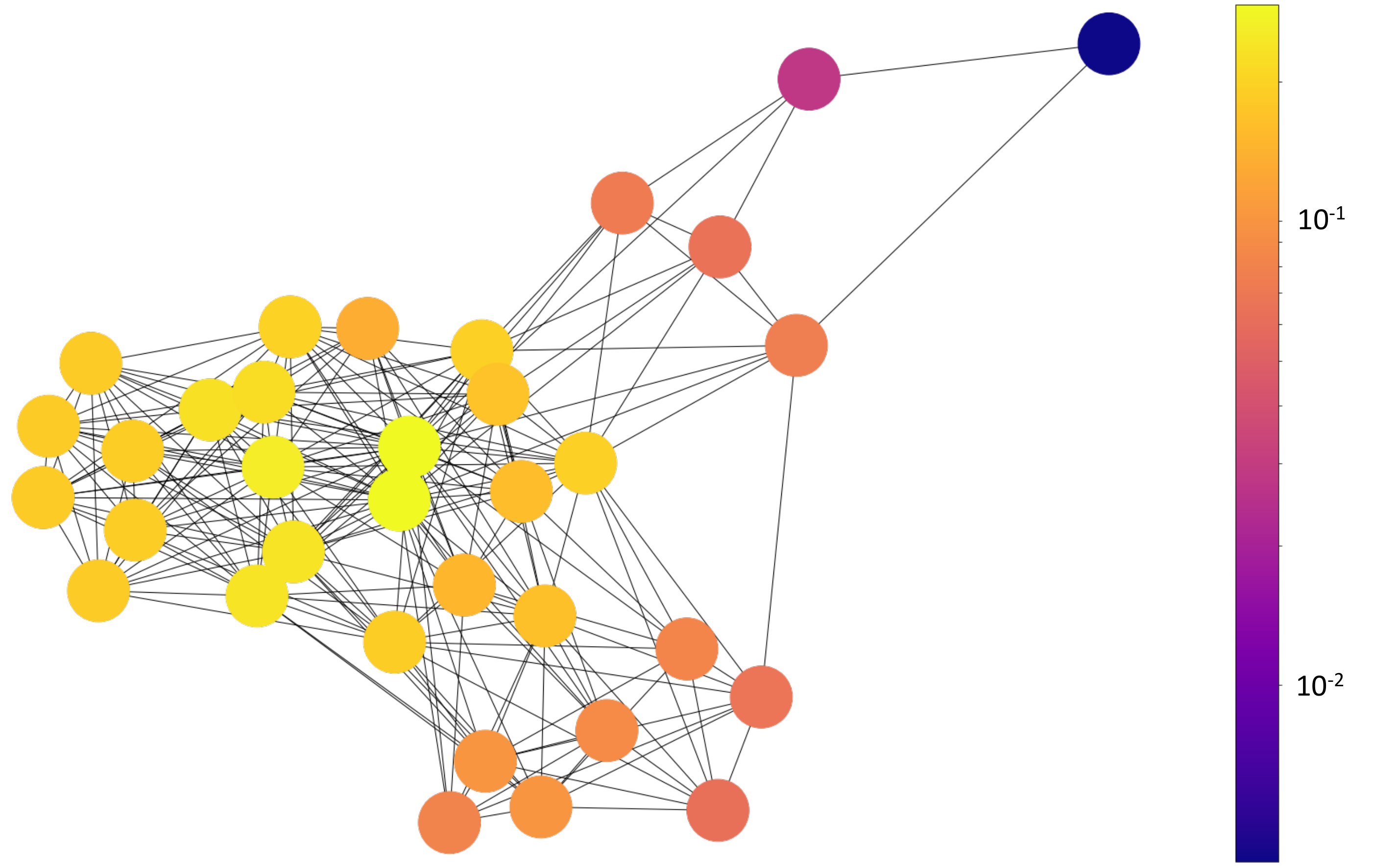}\\
   \end{center}
    \footnotesize{(b)\hspace{5.5cm} (c)}\\
	\includegraphics[scale=0.45]{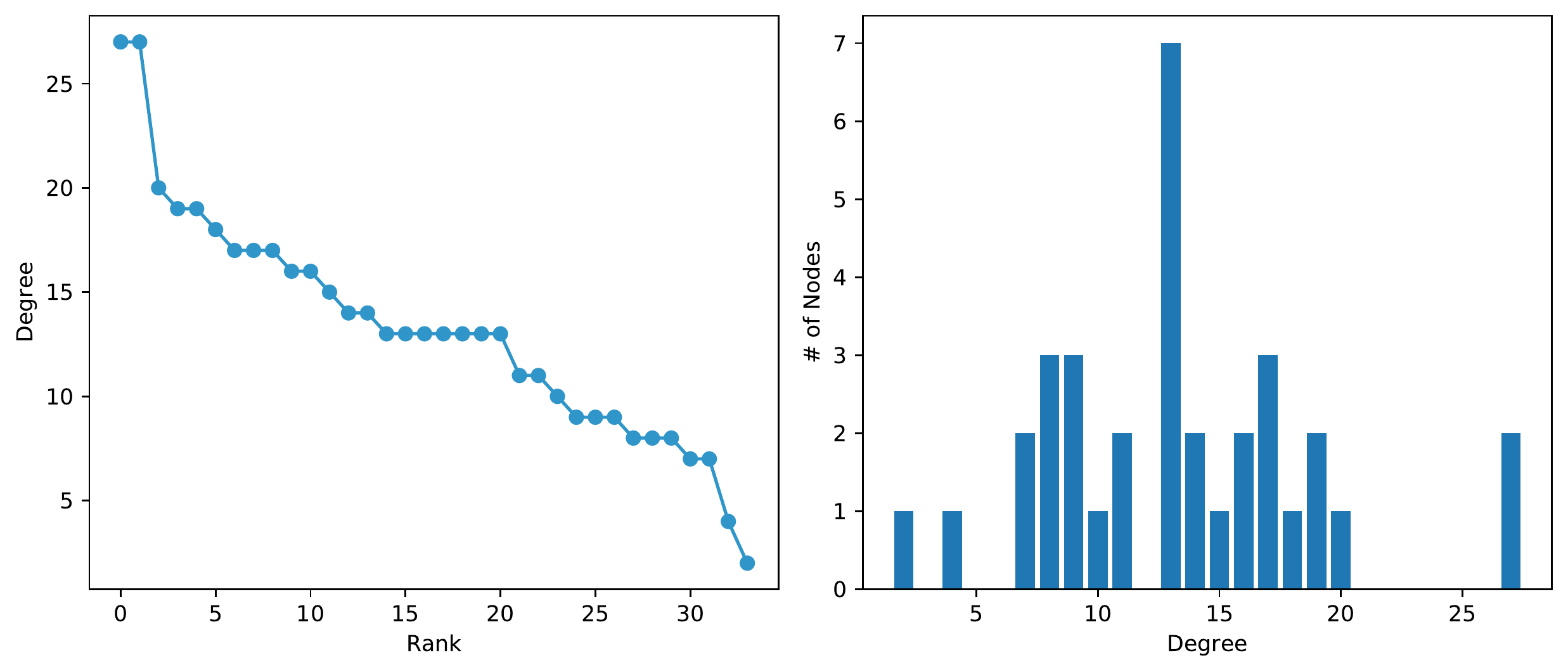}
	\caption{Network Analysis. Panel (a) presents the eigenvector centrality; Panel (b) shows the degree Rank, and Panel (c) is the degree histogram.}
	\label{fig:degree}
\end{figure}

Figure \ref{fig:degree} shows key parameters of the human trafficking network in Chiapas. 
For instance, the eigenvector centrality in Figure \ref{fig:degree}(a) highlights the influence of a node in a network. Relative scores are assigned to all nodes in the network based on the concept that connections to high-scoring nodes contribute more to the score of the node in question than equal connections to low-scoring nodes. 
The exploiter 1, $Ex_1$, 
the participant, $P_1$, and the recruiter/victim, $Rv_1$, have the highest
eigenvector score, thus these actors are connected to many others.
Therefore, these three actors are fundamental for criminal organization.
On the other hand, the network degree in Figure \ref{fig:degree}(b,c) underlines the number of edges that are incident to the node. 
Thus, most of the actors in the criminal network interact with many of the other actors. However, there are 34 actors in the network, each actor may have up to 33 links, therefore, the average degree of centrality about 13.235 is low, which means that the actors are not fully connected. In fact, the criminal network is moderately centralized (44.32\%) and with medium density (0.401), see Table \ref{tab:metrics}. This means that the network may have minimal cohesiveness, and members may find it difficult to share information, money, or products among themselves. Reassuring this, the diameter of the network is three, therefore, isolating a few important nodes should be enough to dismantle the network. 

\begin{figure}[ht]
\includegraphics[scale=0.13]{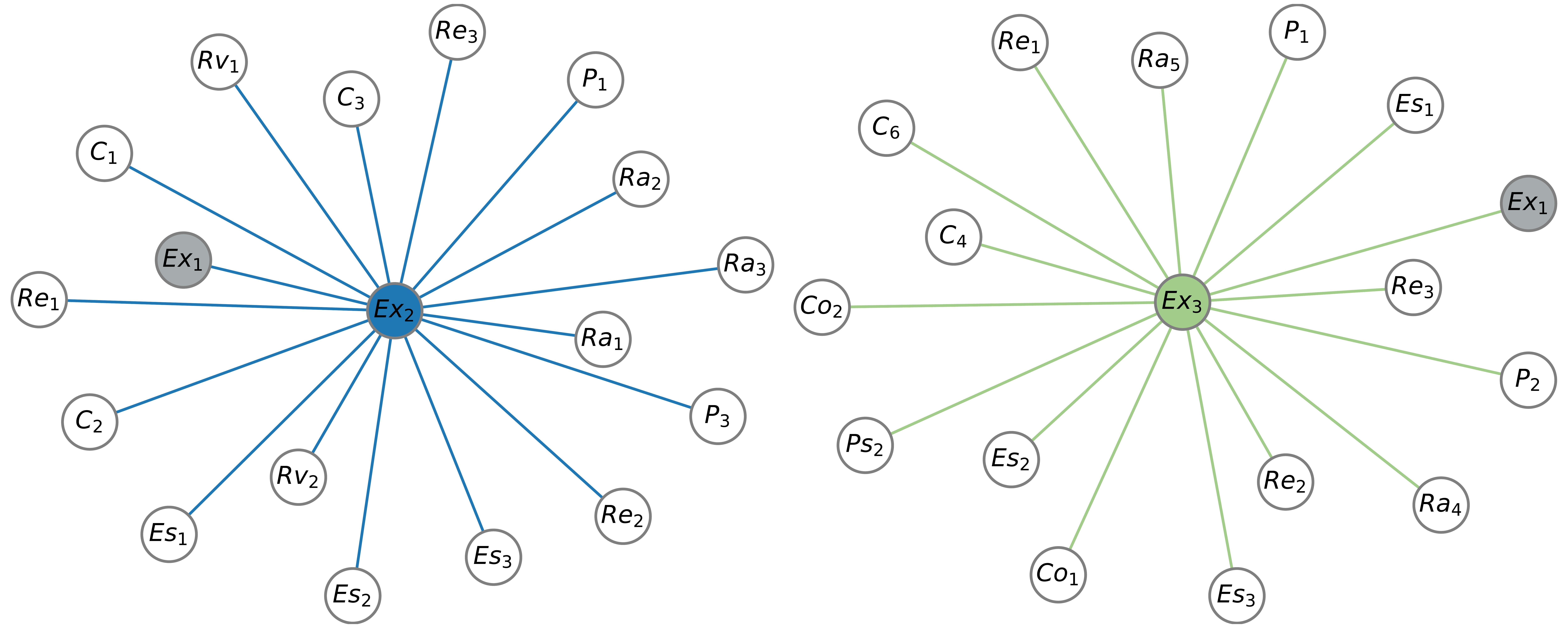}
\caption{Links of Exploiters 2 and 3.}
\label{fig:exploiters}
\end{figure}

\begin{table}[ht]
	\centering
	\begin{tabular}{|c|c|}
		\hline
		\textbf{Topological metric} & \textbf{Score} \\ \hline
		Betweenness centrality    & 2\%           \\ \hline
		Average clustering coefficient    & 0.647           \\ \hline
		Density                    & 0.401            \\ \hline
		Diameter           & 3                \\ \hline
		Average degree                 & 13.235                   \\ \hline
	\end{tabular}
	\caption{Network metrics.}
	\label{tab:metrics}
\end{table}
The cost of removing an actor is set to the degree of the node that represents all the strategies since this reflects the actor's value in the network in terms of its connections. Because of the intricacy of the measurement, each node does not have a monetary cost assigned to it, but the aim of indicating which node is more costly is served by establishing that the more connections there are, the greater the isolation cost.
Exploiters play a central role in the criminal network, in particular, Figure \ref{fig:exploiters}  highlights that exploiter 1 is in contact with exploiters 2 and 3.
In fact, from the collection of data, it is possible to differentiate that exploiter 1 leads the sexual exploitation, whereas also the exploiter 3 is involved. 
The exploiter 2 is leading the laboral exploitation network.

\section{Dismantling Strategies}
A key goal of criminal networks is to craft strategies to destabilize or disrupt illegal organizations \cite{Everton2012}. Thus, this study evaluates how to defragment the human trafficking network and assessed the fragmentation.
\subsection{Removing hubs}
Two individuals have been discovered who function both as \textit{hubs} and \textit{brokers} at the same time; they are exploiter and participant, respectively. After removing the actors Ex1 and Pa1,  Table \ref{tab:met-sin} displays the network's most relevant metrics.
\begin{table}[ht]{\tiny
	\resizebox{\textwidth}{!}{%
		\begin{tabular}{|c|c|c|}
			\hline
			\textbf{Topological metric}        & \textbf{Original score} & \textbf{Score after removing hubs} \\ \hline
			Betweenness centrality   & 2\%                           & 11.44\%                       \\ \hline
			Average clustering coefficient & 0.647                           & 0.606                     \\ \hline
			Density                          &0.401                             & 0.347                     \\ \hline
			Diameter                           & 3                                 & 4                         \\ \hline
			Fragmentation                    & -                            & 0.653                          \\ \hline
			Sverage degree                        &  13.235                           & 10.75                     \\ \hline
		\end{tabular}%
	}}
	\caption{Metrics by isolating \textit{hubs} and \textit{brokers} in the network.}
	\label{tab:met-sin}
\end{table}
Actors are less linked as fragmentation has risen and centrality measurements have fallen. By fragmenting the network, the flow of the three elements that are transmitted across the network is broken. The exploiter offers exploited individuals and information and receives money and information, while the participant contributes information and receives money. Thus, it seems at first instance to remove the hubs, however, this may have several implications as they may have an important connection with public servants and police.

\subsection{Generalized Network Dismantling}

The Generalized Network Dismantling (GND) method belongs to a set of methods called spectrals and locates
the set of nodes to be removed minimizing the costs under the established restrictions \cite{Ren2019}.  
It employs the Fielder vector, which is the eigenvector corresponding to the second smallest eigenvalue of a graph Laplacian matrix. 
This plays a key role in spectral graph theory for graph bi-partitioning. and envelope reduction.
The GND algorithm \cite{Ren2019} incorporates into the optimizing problem that presents the bisection of a network a diagonal matrix with entries with weights that are relative to the degrees of each node.
Thus, the algorithm takes the whole network $G$, with adjacency matrix $A$, and a subnetwork $G^*$, with adjacency matrix $A^*$, as inputs then calculates the degree of each node $i$ in $G$ and in $G^*$, denoted as $k_i$ and $h_i$ respectively, and selects the one with the highest degree with respect to its degree in $G^*$ to add it to the set of nodes to remove, named $R$. 
The process is repeated as many times as needed as long as the set of edges, $E^*$, is not empty. 
The pseudocode of Weighted Vertex Coverage algorithm (WVC) is shown below:
\\
\begin{algorithm}[H]
	\SetAlgoLined
	WVC($G^*$,$G$)\\
	$R = \emptyset$\\
	$k_i= \sum_j A_{i,j}^*$\\
	$h_i= \sum_j A_{i,j}$\\
	\While{$E^* \neq \emptyset$}{
		\hspace{0.5cm} Let $x_i$ be the node with maximum value for $\frac{k_i}{h_i}$\\
		\hspace{0.5cm} $ R\gets R\bigcup\{x_i\}$\\
		\hspace{0.5cm} $ G^*\gets G^*\setminus x_i $\\
		\hspace{0.5cm} $ G\gets G \setminus x_i$\\
		\hspace{0.5cm} upgrade $d$ y $h$
	}
	\Return $R$
	\caption{Weighted Vertex Coverage, for further details see \cite{Bar-Yehuda1981}.}
	\label{algorithm1}
\end{algorithm}
\vspace{1cm}
Consequently, GND generates a matrix $B$, which is referred to as a ``weighted adjacency matrix", from an adjacency matrix $A$ and a removal cost matrix $W$. From the previous matrix, create the diagonal matrix $D_B$ that has as inputs the weighted nodes. Based on the latter, the network's weighted Laplacian matrix is formed $L$, from which the Fielder vector is determined, $\vec{v_2}$, and the network is bisected into the two subnetworks $M$ and $\Bar{M}$. The Weighted Vertex Coverage technique is then used to choose the nodes to be deleted in the following stage. The procedure is then done as many times as indicated. The GND algorithm's pseudocode is as follows:
\\
\begin{algorithm}[H]
	\SetAlgoLined
	GND(A,W)\\
	REM = $\emptyset$ \\
	\While{$CM > C$}{
		\hspace{0.5cm} $B = AW + WA - A$\\
		\hspace{0.5cm} $D_{B_{i,i}} = \sum_{j = 1}^{n} B_{i,j}$\\
		\hspace{0.5cm} $L=D_B-B$\\
		\hspace{0.5cm} Get eigenvalues and eigenvectors of $L$\\
		\hspace{0.5cm} $\vec{v_2}$ $\rightarrow$ $M, \Bar{M}$\\
		\hspace{0.5cm} $M,\Bar{M}$ $\rightarrow$ $G^*=(V^*,E^*)$ bisect the graph\\ 
		\hspace{0.5cm} $R = CVP(G^*,G)$\\
		\hspace{0.5cm} $G = G \setminus R$\\
		\hspace{0.5cm} Append $R$ to REM (set of nodes to remove)\\
		\hspace{0.5cm} upgrade $CM$, $A$ y $W$}
	\Return $Rem$
	\caption{Generalized Network Dismantling \cite{Ren2019}.}
		\label{algorithm2}
\end{algorithm}
\vspace{1cm}
The GND algorithm underlines to first remove \textit{Participant 3} which has $15$ connections. 
This makes the cost close to the average degree of the original network and is a crucial actor for communication between others. 
The second node that the algorithm removes is the \textit{Raitero4}, which has connections to eleven other nodes. 
The algorithm's strategy seems contrary to that of attacking \textit{hubs}, since it attacks actors with a degree close to the average first until it reaches the most connected nodes. 

\subsection{Comparison of Dismantling Strategies}
Here, dismantling strategies previously discussed will be simulated to evaluate the best option for isolating actors and thus dismantling the human trafficking network in Chiapas.
Furthermore, a non-orthodox approach but commonly used in practice is to remove nodes randomly. 
To this end, the removal of each actor from the network is simulated at random. 
\begin{figure}[ht]
	\begin{center}
		\footnotesize{(a)\hspace{5.8cm} (b)}\\[-0.3cm]
		\includegraphics[scale=0.4]{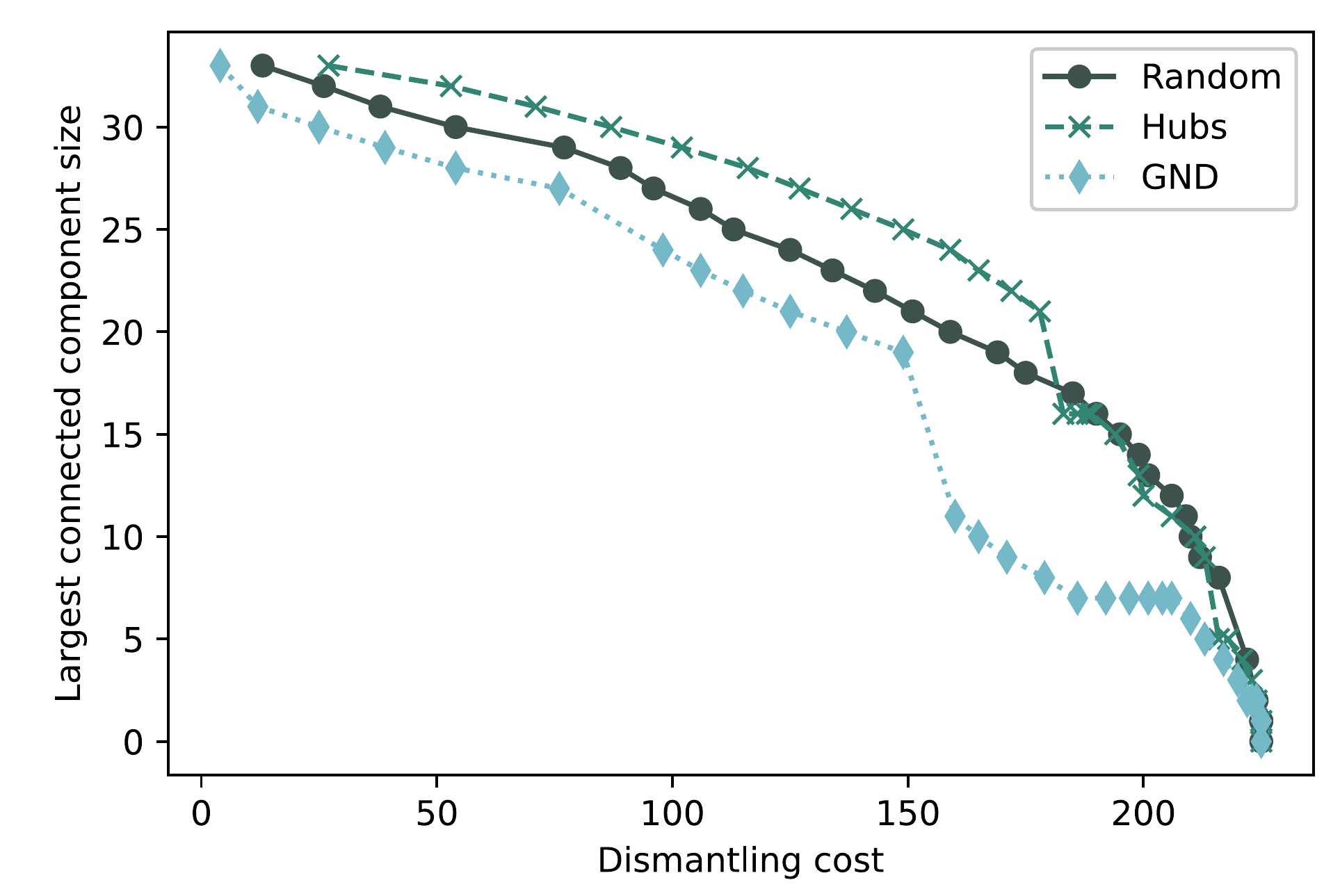}\hspace{-.3cm}
		\includegraphics[scale=0.4]{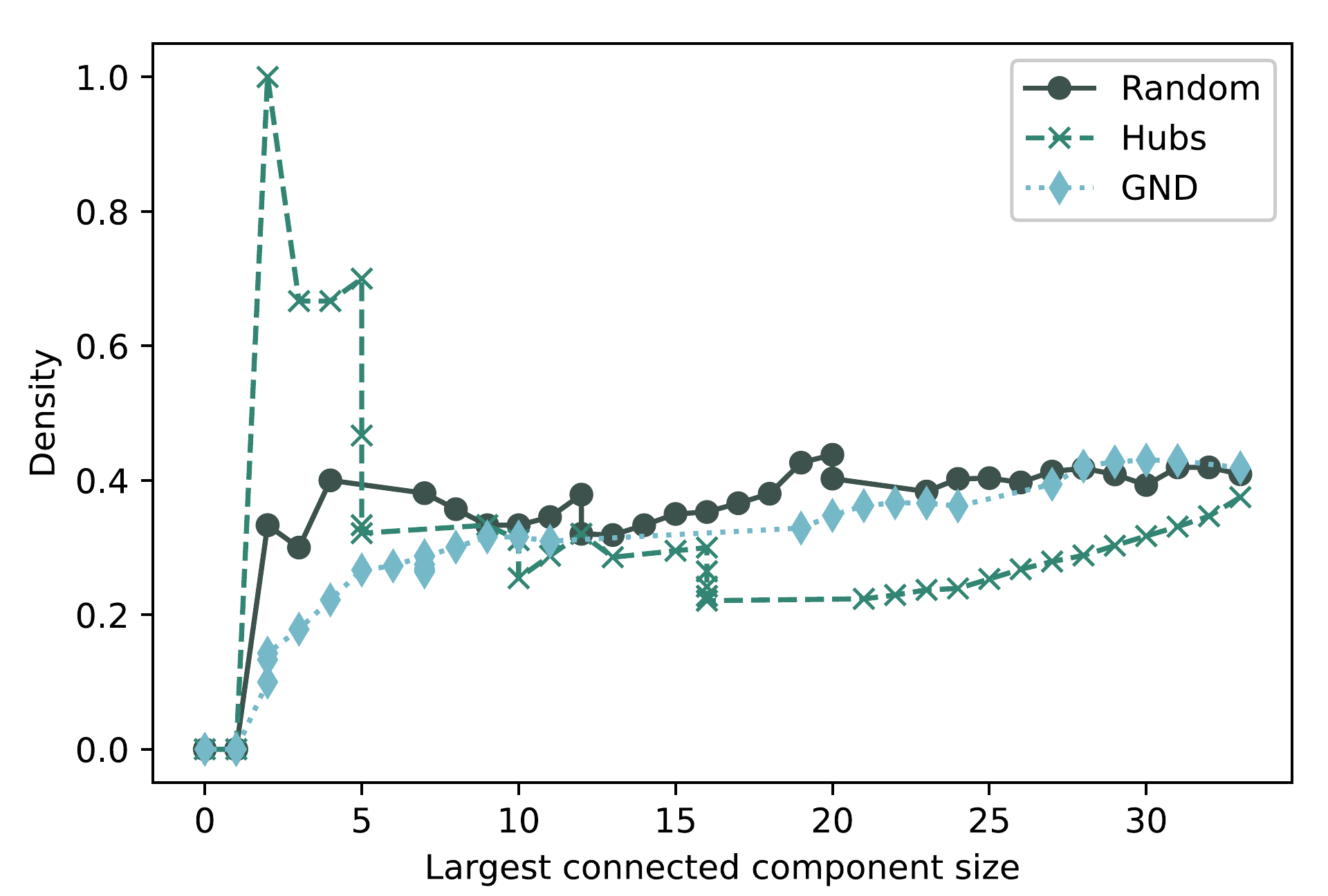}\\
		\footnotesize{(c)}\\[-0.3cm]
		\includegraphics[scale=0.4]{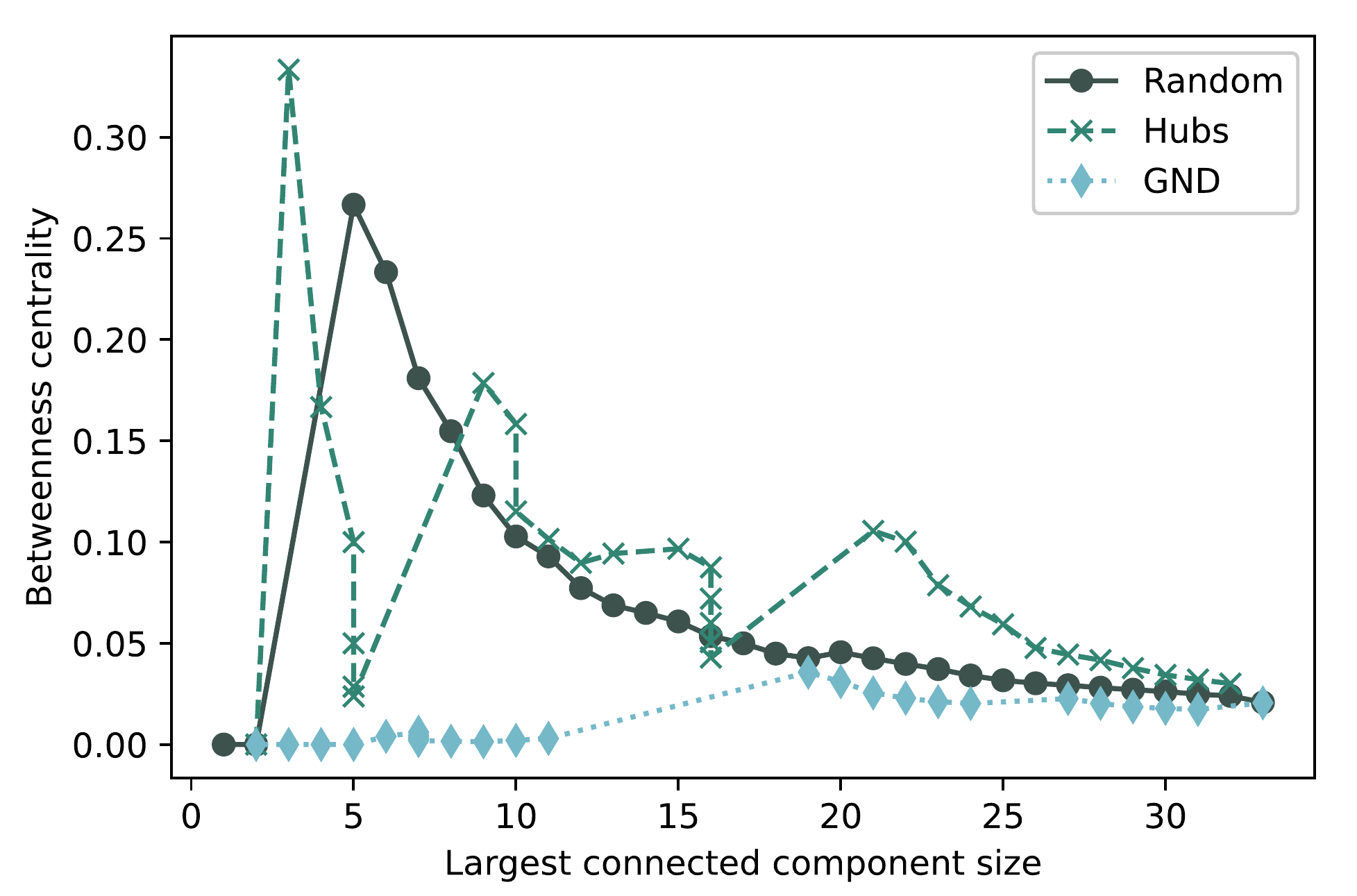}
		\caption{Dismantling strategies performance. Panel (a) presents the cost based on the largest connected component. Panel (b) shows the density as a function of the largest connected component. Panel (c) presents the between centrality as a function of the largest connected component.}
		\label{fig:distmantling}
	\end{center}
\end{figure}

Figure \ref{fig:distmantling}(a) shows the results of the decommissioning cost for the three different dismantling strategies. 
The GND algorithm shows the best results than the other two since it reduces costs and increases network fragmentation.
Figure \ref{fig:distmantling}b it can be seen that only the GND algorithm strategy eliminates the density of the network. The centrality measures also remain lower with the GND algorithm than with the others, an example of this is the betweenness centrality that can be seen in Figure \ref{fig:distmantling}c.

Table \ref{tab:percent} summarizes the cost of dismantling $80\%$, $50\%$, and $20\%$  of the network for each of the algorithms and, once again, the strategy designed with the GND algorithm is the most efficient with a cost of only $186$ for $80\%$.

\begin{table}[ht]
	\resizebox{\textwidth}{!}{%
		\begin{tabular}{|c|c|c|c|}
			\hline
			\textbf{Method}         & \textbf{Dismantling cost 80\%} & \textbf{Dismantling cost 50\%} & \textbf{Dismantling cost 20\%} \\ \hline
			Random isolation      & 214                                & 169                                & 80                                 \\ \hline
			Isolation by \textit{hubs}         & 218                                & 199                                & 127                                \\ \hline
			Isolation based on GND & 186                                & 160                                & 76                                 \\ \hline
		\end{tabular}%
	}
	\caption{Network Dismantling at 20\%, 50\% and 80\%.}
	\label{tab:percent}
\end{table}

\section{Discussion}
Understanding the current problem of human trafficking is as difficult as preventing it. Although the current work focuses on a single microscopic cell in the body of a horrible crime, the breakdown of even one cell means the liberation of tens or perhaps thousands of lives. The study discovered that isolating offenders with a low number of links is the best technique for dismantling a human trafficking network in Chiapas. By lowering isolation costs and maximizing network fragmentation, the General Network Dismantling (GND) algorithm is the optimum option for dismantling a network.

Attacking persons with numerous ties in a criminal network is claimed to be the greatest approach to destroy it since it leaves the rest of the actors in the network without communication,  \cite{Ballester2006}; yet, because networks are incredibly flexible a person can arise to replace it in a short time, \cite{Phillips2015}. Furthermore, the costs of removing leaders are considerable, and most of the strategies employed in the literature do not account for them. As a result, targeting those with a lot of connections is only a temporary solution, \cite{Everton2012}. 
Attacking these individuals merely makes the network more concealed and tougher to destroy in the case of decentralized networks that lack a leader or a few leaders, \cite{Brafman2006}.

In the first five rounds of both the approach against hubs and the GND strategy, the exploiter actor, with the associated number 1, is isolated. This is intriguing since he appears to be a key figure in dismantling the network, and his involvement in establishing contact with key victims and criminals is crucial. In future works, a weight associated with the role played by criminals could be included so that the isolation costs are more closely related to reality.

The main contribution of this work is the network built with real data in an area that is little analyzed for a crime that has received little attention. Human trafficking networks have not been studied by the international community, academics, or governments in general. Human trafficking, unlike drug or arms trafficking, has been overlooked; yet, this work is believed to be one of the first to provide efforts on the research of criminal networks of this sort, setting a precedent on their topology.

Dismantling criminal organizations that exploit individuals will not put a stop to the crime since new participants will constantly emerge to resurrect the business to meet the current demand. Human trafficking networks must be dismantled in tandem with measures that reduce demand for trafficked persons, such as raising awareness among young people, broadening alternatives for those most likely to join criminal groups, and improving the design and functioning of public institutions.

To isolate a node from the network, the procedure entails more than simply recognizing it as a member of the network; it must also be arrested, charged, found guilty, and punished such that it has a significant impact on the network's structure, \cite{Duijn2015}. All of this comes at a cost to the state; but, when considering the long-term implications of human trafficking, the expenses of dismantling will never be comparable to what the victims must pay for the rest of their life.

\bibliographystyle{unsrt}

\bibliography{mybibfile}

\end{document}